\documentclass[11pt]{article}
\usepackage[utf8]{inputenc}
\usepackage{xcolor}
\usepackage{graphicx}
\usepackage[left=2.0cm, right=2.0cm, top=2.0cm, bottom=2.0cm]{geometry}
\usepackage{apacite}
\usepackage{booktabs}
\usepackage{amsmath}
\usepackage{caption}
\usepackage{float}
\usepackage{enumitem}
\usepackage{tabularx}

\AtBeginDocument{}

\title{\Large \textbf{PLACE AND POLITICAL BEHAVIOR ONLINE \\ DURING THE LOS ANGELES BLAZE}}
\author{\small Sofi Koponen \footnotemark[1] \and \small Ali Salloum \footnotemark[1] \and \small Luca Maria Aiello \footnotemark[2] \and \small Mikko Kivelä \footnotemark[1] \and \small Arttu Malkamäki\footnotemark[1]}
\date{}

\begin{document}

\maketitle 
\thispagestyle{empty}
\footnotetext[1]{\footnotesize Aalto University, Finland}
\footnotetext[2]{\footnotesize IT University of Copenhagen, Denmark}

{\let\abstractname\relax 
\begin{abstract}
\noindent Understanding how online communication during extreme weather crises enables or constrains climate mobilization is critical. During a crisis, the online world connects distant audiences and complicates the relationship between place and political behavior. We analyze all sociopolitical content from Bluesky about the January 2025 Los Angeles wildfire event, reporting how participation in collective action and allocation of attention vary by geographic distance. Participation varies nonlinearly with proximity, with local and international users participating more than regional and national users. Attention concentrates around regional and national users engaging primarily in viral, non-participatory political discussion. Analysis reveals a geographic mismatch between participation and attention, illuminating how online communication concentrates attention away from collective action amid an exacerbating climate crisis. \\
\\
\noindent \textit{climate change}; \textit{connective action}; \textit{context collapse}; \textit{political communication}; \textit{social media}
\end{abstract}}

\section*{\large INTRODUCTION}

\noindent \textit{All politics is local} encapsulates Tip O'Neill's conviction, as former Speaker of the US House of Representatives, that local concerns and community interests define political behavior, even amid national controversies \cite{oneill_all_1994}. However, political communication in the online age traverses geographic and social contexts, connecting previously separate audiences and complicating the relationship between place and behavior \cite{davis_context_2014}. Frames and ideas circulate rapidly between contexts, reinforcing or transforming one another through recursive feedback \cite{padgett_emergence_2012}. If online communication moderates the effect of place on behavior, could the digital world also dilute our ability to act collectively in the face of major environmental crises?

Understanding the spatial and structural distribution of attention and mobilization matters for assessing how online communication constrains and enables political action for the environment \cite{cologna_extreme_2025, oboyle_psychological_2026}. Here, we analyze all data from the online platform Bluesky about the January 2025 wildfires in Los Angeles, CA. The event claimed dozens of lives and caused \$76–131 billion in losses, alongside economic, social, and health impacts \cite{li_economic_2025}. It also ignited a political debate, including on emergency response, political responsibility, and climate change \cite{amin_ecolinguistic_2026}. We focus on documenting how online communication across audiences at varying geographic distances conditioned i) the extent of participation in collective action and ii) the allocation of presumably scarce attentional capacity.

First, we annotated posts by sociopolitical genre (from political commentary to causal attribution) and users by geographic locality (from local to international). Second, we coded posts for the presence or absence of collective action participation and, where present, the corresponding mode of engagement (from problem identification to action execution). Third, we fitted a nested stochastic block model to a bipartite user-to-post-by-repost network to infer groups of users with similar connectivity patterns, describe their composition, and trace their attention allocation.

We show that both local and international users are more likely to participate. Regional and national users garner more attention, yet engage primarily in viral, non-participatory political commentary. While Bluesky's left-leaning user base limits cross-platform generalizability \cite{quelle_bluesky_2025}, that relative ideological homogeneity provided a useful baseline for extending research on geographic variation in online communication by demonstrating a nonlinear association between place and online political behavior \cite{jaidka_estimating_2020}.

\section*{\large MATERIALS AND METHODS}

We collected all communication available through Bluesky's firehose between December 17, 2024, and May 1, 2025. First, we used an open-source large language model (LLM), to identify event-related content, yielding 36,073 original posts and corresponding reposts. Second, we applied to cluster the posts into semantic microtopics. Third, we manually aggregated the resulting microtopics into eight issues and developed a codebook with comprehensive descriptions of each. We provided the codebook to a zero-shot LLM to assign each post to the corresponding issue. However, since only four issues---political commentary, socioeconomic concern, solidarity expression, and causal attribution---provided a meaningful basis for coding collective action, we focused our analysis on them, retaining 17,119 posts.



We classified the remaining posts for participation in collective action using a zero-shot LLM classifier in two stages. Following a recent framework \cite{pera_extracting_2025}, we first assigned each post a binary participation label and then classified all posts exhibiting participation by the mode of engagement, whether problem or solution identification, call-to-action, intention, or execution.

To infer users' localities at a meaningful resolution, we extracted location signals from their profiles and recent posts using speaker-anchored lexical patterns and named-entity recognition. An LLM classified each user into one of four categories: local (LA), regional (CA), national (US), or international (non-US). We recovered location for 54\% of users.

Lastly, we constructed a bipartite user-post network comprising authoring and reposting edges, with 121,175 users, 17,119 posts, and 341,755 edges. We inferred the network's latent block structure by fitting a Bayesian hierarchical, degree-corrected stochastic block model to the network \cite{peixoto_hierarchical_2014}. To mitigate overfitting and identify a well-fitting model, we selected the solution with the minimum description length across 100 random initializations. We focused on the second level of the hierarchy, where users with statistically similar connectivity patterns belong to the same block. 

\section*{\large RESULTS}

\subsection*{Collective action participation}

\begin{figure}[t!]
    \centering
    \includegraphics[width=\linewidth]{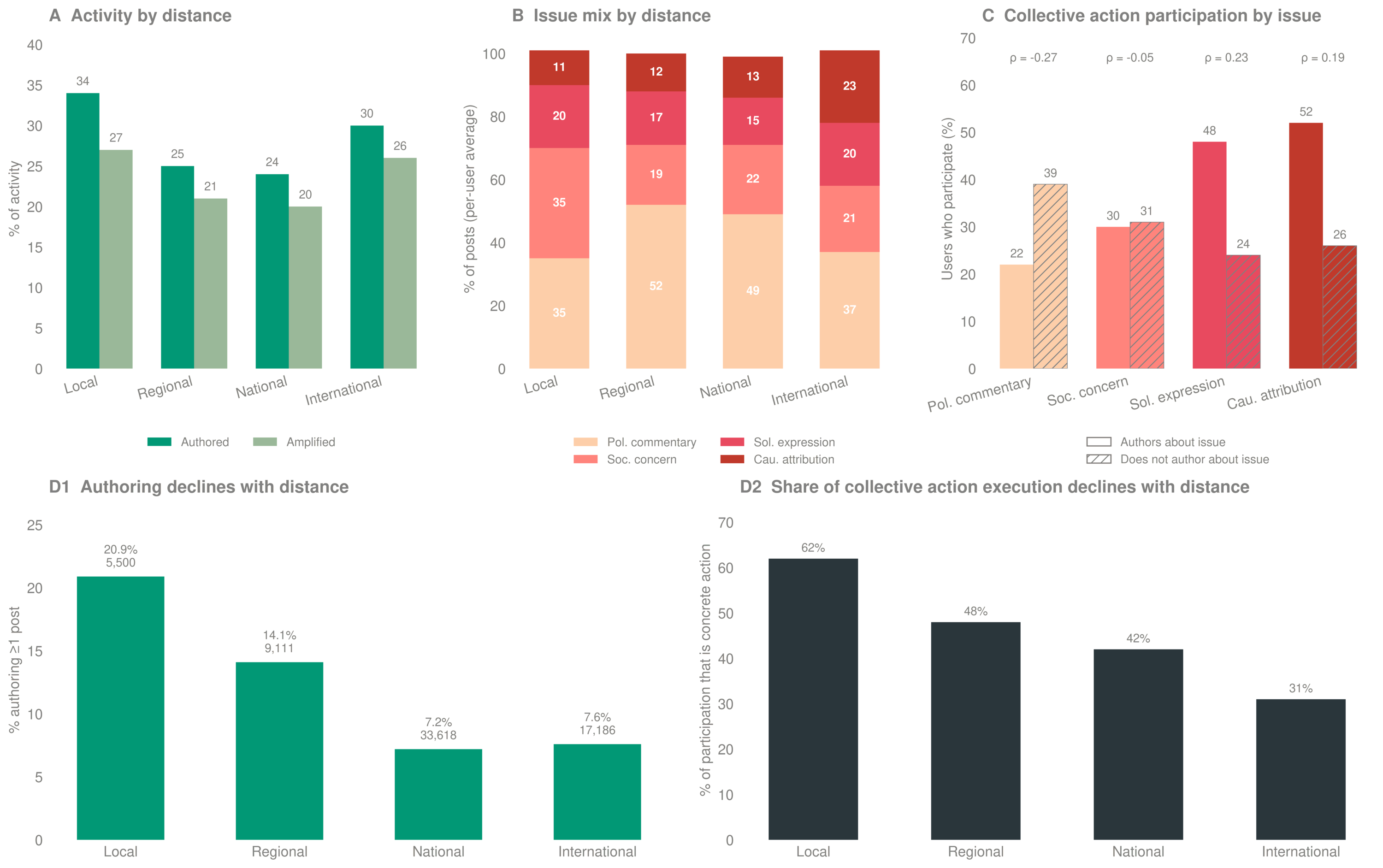}
    \caption{\small \textbf{Collective action participation by issue and locality.} A) Share of authored posts and reposts containing participation content within each locality group. B) Issue composition by locality group. C) Share of users posting participation content among those who post about each issue and those who do not; values above bars report Spearman $\rho$. D1) Share of users authoring at least one post, with group sizes. D2) Share of participation content that classifies as concrete action.}
    \label{fig:locality_stat}
\end{figure}

Participation content is unevenly distributed across locality groups (Fig.~\ref{fig:locality_stat}A). Its share of authored posts is highest among local and international users (34\% and 30\%, respectively) and lowest among regional and national users (25\% and 24\%). This U-shaped pattern also appears in reposts, where the corresponding shares are 27\%, 21\%, 20\%, and 26\%. In every locality group, participation content accounts for a smaller share of reposts than of authored posts.


Issue composition varies with locality (Fig.~\ref{fig:locality_stat}B). Political commentary accounts for approximately half of regional and national users' posts (52\% and 49\%), compared with 35\% among local users and 37\% among international users. Their concentration therefore follows an inverted U-shaped pattern, meaning that the groups with the largest shares of political commentary also have the smallest shares of participation content.


At the user level, posting political commentary is negatively associated with posting participation content (Fig.~\ref{fig:locality_stat}C). Among users who post political commentary, 22\% also post participation content, compared with 39\% among those who do not. The corresponding Spearman correlation is $\rho = -0.27$. Solidarity expression and causal attributions show the opposite association: participation content is posted by 48\% versus 24\% of users who do and do not post in the solidarity expression genre, and by 52\% versus 26\% of users who do and do not post causal attributions. Their correlations are $\rho = 0.23$ and $\rho = 0.19$, respectively. Socioeconomic concerns show little association with participation posting (30\% versus 31\%; $\rho = -0.05$).

These comparisons are consistent with a relationship between issue composition and the lower participation shares observed among regional and national users. They do not, however, establish that political commentary causes lower participation or explain the differences between locality groups.


The remaining panels distinguish authoring from the form of participation. The share of users authoring at least one post falls from 20.9\% among local users to 14.1\% among regional users and 7.2\% among national users, before increasing slightly to 7.6\% among international users (Fig.~\ref{fig:locality_stat}D1). By contrast, the share of participation content classified as \textit{execution} declines monotonically with distance: 62\% for local users, 48\% for regional users, 42\% for national users, and 31\% for international users (Fig.~\ref{fig:locality_stat}D2).


The results distinguish the prevalence of participation content from its action component. Participation content is most prevalent at the local and international extremes, whereas its action component is greatest locally and decreases with distance. Regional and national users concentrate more heavily on political commentary, an issue negatively associated with participation posting at the user level. Thus, distance and issue composition reveal different patterns of engagement, where participation content does not decline uniformly with distance, even though the share classified as concrete action does.


\subsection*{Attention allocation}

\begin{figure}[t!]
    \centering
    \includegraphics[width=\linewidth,trim={100 30 100 30},clip]{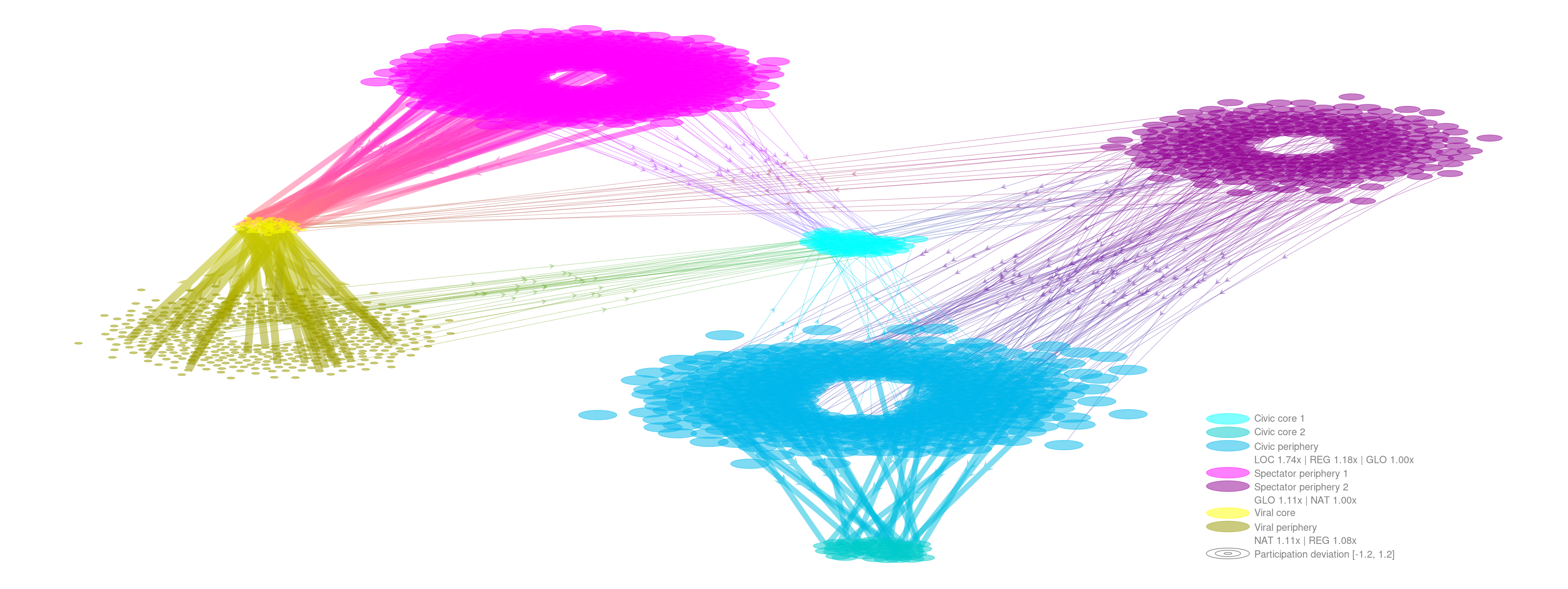}
    \caption{\small \textbf{Attention allocation by block strata.} Vertices are arranged into blocks identified by a Bayesian stochastic block model and grouped into \textit{civic}, \textit{spectator}, and \textit{viral} strata. Vertex colors distinguish blocks and their core or periphery roles. Vertex size represents participation relative to the block average, combining authored and amplified participation weighted by the block's posting and reposting activity. Directed edges represent reposting; edge width reflects amplification strength, measured as repost volume per user. The edge color gradient indicates direction. The legend reports locality multipliers for each stratum.}
    \label{fig:block_graph}
\end{figure}

Aggregating the block model output yields three strata that differ sharply in attention received relative to content produced (Fig.~\ref{fig:block_graph}). The \textit{viral} stratum holds 29\% of users and authors 29\% of posts, yet receives 49\% of all reposts, most of them captured by a small core (2.5\% of users, 46\% of reposts) amplified by a large periphery. It is the most political stratum (67\% authored, 71\% amplified) and the least participatory (16\% authored, 11\% amplified). The \textit{civic} stratum, in turn, is its counterpart: with 12\% of users it authors 25\% of posts, is the least political (37\% authored, 41\% amplified) and amplifies the most participatory content (30\%), yet receives only 19\% of total reposts. The remaining \textit{spectator} stratum holds 60\% of users, authors 46\% of posts and receives 33\% of reposts; it authors the most participatory content (27\%) but amplifies more political (57\%) and less participatory (24\%) content than it produces.

Attention thus concentrates on the least participatory, most political region, largely through self-reinforcement. The viral hub makes 62\% of all reposts and supplies 81\% of its own authors' attention, while directing only 9\% to the civic stratum. The civic source instead relies on itself (46\% of its attention, 2.6$\times$ baseline) and the spectator stratum (25\%, 1.2$\times$). The attention of the viral stratum is also extremely skewed: its top 1\% of authors capture 75\% of its reposts and are less participatory than its typical author (9\% vs.\ 16\%). In civic stratum the top 1\% capture only 39\% and remain about twice as participatory as the viral stratum's core tier (17\%). Political content thus reaches viral levels, sustained by a large base of active amplifiers, while participatory content lacks such a base, circulating within a smaller group and at the networks edges, at a far lower scale, spread more evenly across authors. 

The strata also amplify different authors. The viral hub favors regional authors (1.19$\times$ baseline) and amplifies national authors at roughly baseline (1.03$\times$). In sum, regional and national authors receive 83\% of the hub's reposts (among authors with a known location). The civic stratum instead over-amplifies local (2.06$\times$) and international (1.38$\times$) authors. The hub's attention therefore centers on highly political regional and national authors, while the more participatory local and international authors are amplified mainly outside it.

The network structure paints a picture of the discourse, where participation does appear throughout the network, concentrating in spaces over-represented by local and international users, but is overshadowed by viral non-participatory political content authored by regional and national authors. Participatory content is left to circulate in smaller, secondary spaces and at the network's edges, unable to reach the level of attention captured by political content. The viral hub thus inverts the U-shaped pattern of participation (Fig.~\ref{fig:locality_stat}A), concentrating attention on regional and national authors and political content, and away from the more participatory local and international authors.

\section*{\large DISCUSSION}

Our analysis of online communication on Bluesky during the January 2025 blaze in Los Angeles, CA, reveals a nonlinear association between place and political behavior. It therefore extends prior work on geographic variation in online communication \cite{jaidka_estimating_2020}, while also highlighting that social interactions across online and offline domains remain embedded in, and shaped by, those in the ecological domain \cite{bodin_improving_2019}.

Collective action participation content followed a U-shaped pattern across distance, whereas the share classified as action declined with distance. Regional and national users concentrated more heavily on political commentary, which were negatively associated with participation posting. Put another way, the convergence of audiences online connected geographically distinct contexts without producing a common orientation toward participation \cite{davis_context_2014}. Attention and participation also diverged, with high-reach regional and national accounts occupying most central positions in the user-post network, while local and international participation content attracted less visibility. The event brought these social worlds into contact, but their contributions received unequal attention.

Whether the attention--participation divide characterizes environmental crisis discourse more generally remains open. Our findings concern one event on one notably left-leaning platform \cite{quelle_bluesky_2025}, with locality recoverable only for 54\% of users. However, by removing the partisan divide and allowing for meaningful variation in behavior and locality, the population was perfectly suitable for our analysis. Yet, neither participation posting nor network position establishes offline mobilization or institutional influence, warranting further comparisons across events and platforms, linked to mobilization outcomes, to assess how these patterns constrain, enable, and shape environmental action \cite{cologna_extreme_2025, oboyle_psychological_2026}. The analysis nonetheless reveals a potential disconnect. The regional and national audiences that could connect local concerns to institutional responses attracted substantial attention, but their discourse was predominantly non-participatory.





\bibliographystyle{apacite}
\bibliography{reference_list}

\end{document}